\documentclass[a4paper,11pt]{article}
\usepackage{graphicx}  
\usepackage[utf8]{inputenc}
\usepackage{amsmath,amssymb}

\def \sec{\begin{section}}
\def \esec{\end{section}}

\def \om {\omega}

\def \pr {\partial}
\def \PD {\mathcal{D}}
\def \la {\lambda}
\def \La {\Lambda}
\def \qs {\qquad}

\def \l  {\left(}
\def \r  {\right)}

\begin{document}
\begin{flushright}
ITEP-TH-32/12
\end{flushright}

\vspace{1cm}

\begin{center}
{\Large \bf CP(N-1) model on finite interval in the large N limit}
\end{center}

\vspace{0.5cm}

\begin{center}
{ A. Milekhin\footnote{milekhin@itep.ru}} \\ 
\emph{Institute of Theoretical and Experimental Physics,}\\
\emph{Moscow 117218, Russia} \\
\emph{And} \\
\emph{Moscow Institute of Physics and Technology,} \\
\emph{Dolgoprudny 141700, Russia }
\end{center}
\date{}


\begin{abstract}
The $\mathbb{C}$P(N-1) $\sigma$ model on finite interval of length R with Dirichlet boundary conditions is analysed 
in the 1/N expansion. The theory has two phases, separated by a phase transition at $R \sim 1/\La$, $\La$ is dynamical scale
of the $\mathbb{C}$P(N-1) model. The vacuum energy dependence of R, and especially Casimir-type scaling $1/R$, is discussed.

\end{abstract}

\sec{Introduction}
The large N expansion is suitable to study non-perturbative behavior of a variety of models in different physical 
situations(see \cite{JEAN} for a review). Within this technique many important features such as dynamical mass generation,
asymptotic freedom and an absence of spontaneous continuous symmetry breaking in two dimensions could be seen. 

In what follows, we will consider two dimensional non-linear $\mathbb{C}P$(N-1) $\sigma$ model on finite interval 
of length R with Dirichlet boundary conditions, that is, on a ribbon. 
In infinite space it was solved by Witten \cite{WITT} by means of the large N expansion. 
The theory is asymptotically free and possesses dynamical mass generation 
via dimensional transmutation:
\begin{equation}
\label{eq:dyn_scale}
\La^2=\La_{uv}^2 \exp \l \frac{-4 \pi}{g^2} \r
\end{equation}
where $\La$ is dynamical scale, $\La_{uv}$ is ultraviolet cutoff and $g$ is a bare coupling constant. It is well-known, 
that the $\mathbb{C}P(N-1)$ model is the effective low-energy theory on a non-Abelian string worldsheet(\cite{STR_BB}). 
Therefore, such a geometry with two Dirichlet boundary conditions can be thought as a non-Abelian string between two 
branes.

In this article we will obtain the following results. The theory has non-trivial R dependence: 
at $R>>1/\La$ it is in the "confining phase"
and the mass gap is present, at $R<<1/\La$ it is in  the "Higgs phase" and there is no mass gap. 
Very similar behavior occurs in
"twisted mass" deformed $\mathbb{C}$P(N-1) model, where "twisted mass" parameter plays the role of R(see \cite{DEF_CPN}
from where the names of the phases were taken). Despite the existence of the mass gap, the vacuum energy has Casimir-type
behavior $1/R$. We will discuss it in the light of the works \cite{SHIF}, \cite{CAS_QCD}.

\esec

\sec{Gap equation}
The considerations below are very similar to those in \cite{DEF_CPN}. We start with the action
\begin{equation}
\mathcal{L}=\frac{N}{g^2}(\pr_\mu-iA_\mu)n_i(\pr^\mu+iA^\mu)n^{*i}-\lambda(n_i^*n^i-1)+\
\frac{\theta}{2 \pi}\epsilon_{\mu \nu} \pr^\mu A^\nu
\end{equation}
Where $\la$ and $A_\mu$ are Lagrange multipliers. $\la$ impose the constraint $n_i^* n^i=1$, $A_\mu$ are just a dummy fields
which could be eliminated by equation of motion:$A_\mu = i n_i^* \pr_\mu n^i $ but make U(1) invariance obvious.
All the fields live on finite interval of length R with Dirichlet boundary conditions:
\begin{equation}
n^1(0)=n^1(R)=1\ ;\ n^i(0)=n^i(R)=0,\  \ i=2,..,N \label{eq:BOC}
\end{equation}
Note that this boundary conditions break translation invariance.

To solve the theory in the large N limit we should integrate over $n^k$ in path-integral to obtain effective action 
for $\la, A_\mu$. 
\begin{eqnarray}
Z=\int \PD A \PD \la \PD n^i \PD n^{*i} \;  \exp \bigl(i \int d^2x \bigl(  - \
\frac{N}{g^2} n^i (\pr_\mu+iA_\mu)^2 n^{*i} - \nonumber \\
\la (n_i n^{*i}-1) + \frac{\theta}{2 \pi}  \epsilon_{\mu \nu} \pr^\mu A^\nu  \bigr) \bigr)
\end{eqnarray}

It will be useful to separate $n^i$ into $n^1=\sigma$ , (N-1) component $n^i$ and integrate over only the last ones.
After rescaling $n^i$, gaussian integration leads us to
\begin{eqnarray}
Z=\int \PD A \PD \la \;  \exp \bigl(-(N-1) Tr \log(-(\pr_\mu+iA_\mu)^2-m^2) + \nonumber \\
i \int d^2x( (\pr_\mu \sigma)^2 - m^2 \sigma \sigma^* + \frac{Nm^2}{g^2}) + \frac{i \theta}{2 \pi}\
\int d^2x \epsilon_{\mu \nu} \pr^\mu A^\nu \bigr)
\end{eqnarray}
where $m^2=\cfrac{\la g^2}{N}$

Now we will use the steepest descend method with the uniform saddle point: $A_\mu = 0\ ,\ m = const, \sigma=const$ and in
the leading order we can neglect the difference between N and N-1.
Also, thought the translation invariance is broken, it is reasonable to expect that we will describe the behavior correctly
at least at qualitative level. 
Varying action with respect to $m^2, \sigma^*$, we obtain saddle-point equation:
\begin{eqnarray}
\label{eq:begin}
g^2 Tr \frac{1}{(-\pr_\mu)^2 - m^2 + i\epsilon} + i\int{(1 -  \frac{g^2 \sigma^2}{N} )} d^2x  = 0 \\
m^2 \sigma = 0
\end{eqnarray}
The second equation implies that $\sigma=0$ or $m=0$. Let us consider the case $\sigma=0$. Then the first equation 
reads(the trace should be computed with respect to (\ref{eq:BOC})):
\begin{equation}
\label{eq:ir_summ}
i+g^2 \sum_{n=1}^{+\infty} \int_{-\infty}^{+\infty} \frac{dk}{2 \pi R} \frac{1}{k^2 - (\frac{\pi n}{R})^2 -m^2+i \epsilon}=0
\end{equation}

Using the identity:
\begin{equation}
\label{eq:sum}
\sum_{\mathcal{Z}} \frac{1}{(\frac{\pi n}{R})^2 + \om^2}=\frac{2R}{\om} \l \frac{1}{2}+\
\frac{1}{\exp(2R\om)-1} \r
\end{equation}
and after the Wick rotation we arrive at
\begin{eqnarray}
1-\frac{g^2}{2 \pi R} \int_{0}^{+\infty}dk\bigl( \
\cfrac{R}{\sqrt{k^2+m^2}} + \nonumber \\
\cfrac{2R}{\sqrt{k^2+m^2}} \frac{1}{\left(\exp(2R\sqrt{k^2+m^2})-1\right)} - \
\cfrac{1}{k^2+m^2} \bigr)=0 \label{eq:BIG}
\end{eqnarray} 

\esec

\sec{Analysis}
Let $x=1/m $ and
\begin{equation}
Q(\frac{x}{R})=\int_0^{+\infty}\frac{2 dk}{\sqrt{k^2+\cfrac{R^2}{x^2}}}\cfrac{1}{\left(\exp(2\sqrt{k^2+\cfrac{R^2}{x^2}})-1\right)} \label{eq:Q}
\end{equation}
If $\La_{uv}$ is ultraviolet cutoff, (\ref{eq:BIG}) leads to
\begin{equation}
1-\cfrac{g^2}{2 \pi R}\left(R \log(\La_{uv} x) + R Q(x/R) - \cfrac{\pi x}{2} \right) = 0
\end{equation}
It is more convenient to rewrite it as, recalling (\ref{eq:dyn_scale}):
\begin{equation}
\label{eq:GAP}
\cfrac{2 \pi}{g^2}-\log(\La_{uv} R)=-\log(\La R)=\log(x/R) + Q(x/R) - \cfrac{\pi x}{2 R}
\end{equation}
If $x<<R$, Q could be calculated using saddle-point approximation, with $k=0$ as a saddle-point:
\begin{equation}
\label{eq:approx}
Q(x/R) \approx \cfrac{\sqrt{\pi x} e^{-\cfrac{2R}{x}}}{\sqrt{R}},\; x<<R
\end{equation}
so Q is exponentially suppressed and so negligible. In the limit $R \rightarrow +\infty$, $\cfrac{\pi x}{2R}$ is also 
negligible and we repeat Witten's result(\cite{WITT}): 
\begin{equation}
\cfrac{2 \pi}{g^2} = \log(\La_{uv} x_0) 
\end{equation}
It is interesting to find 1/R corrections. If $x_0=1/m_0, \la_0$ are solutions for $R=+\infty$, then trivial 
calculation yields
\begin{equation}
x=x_0+\cfrac{\pi x_0^2}{2R}+\cfrac{3 \pi^2 x_0^3}{8R^2} + O(1/R^3)
\end{equation}
Therefore,
\begin{equation}
\label{eq:m2}
m^2=\cfrac{g^2 \la}{N}=\cfrac{1}{x^2}=\cfrac{1}{x_0^2}-\cfrac{\pi}{x_0 R}+O(1/R^3)
\end{equation}
\begin{equation}
\label{eq:m}
m=m_0-\frac{\pi}{2R}-\frac{\pi^2}{8 m_0 R^2} + O(1/R^3)
\end{equation}

In the next section we will use this expansion to calculate 1/R corrections to vacuum energy.

Another mode is $x>>R$. $Q(+\infty)=+\infty$, because the integral is divergent at lower bound. This mode is much more 
harder to deal with. So we calculated the right side of (\ref{eq:GAP}) numerically. The result is shown in the figure 
below. The blue curve is the right side of (\ref{eq:GAP}), the red one is without $Q(x/R)$. At large $x/R$ it has an
asymptotic value $-1.26$, so 
\begin{equation}
Q(x/R) \approx \frac{\pi x}{2R} - \log(\frac{x}{R}) - 1.26 + ...,\; x >> R
\end{equation}
It is possible to calculate the next order term:
\begin{equation}
\label{eq:asympt}
Q(x/R) = \frac{\pi x}{2R}+\log \l \frac{R}{x} \r -(\log(2 \pi)-\gamma)-\frac{\zeta(3)}{2 \pi^2}\left(\frac{R}{x}\right)^2+
O((R/x)^3)
\end{equation}
where $\gamma \approx 0.577...$ - the Euler–Mascheroni constant.
Recalling that $1/x=m$, 
\begin{equation}
m^2=\frac{2 \pi^2}{R^2 \zeta(3)}\left(\log(\La R) -(\log(2 \pi)-\gamma) \right)
\end{equation}

Note that the gap equation has a solution only for R large enough. 

\begin{figure}
\centering
\includegraphics[trim=0.23in 0.23in 0.23in 0in,scale=0.25]{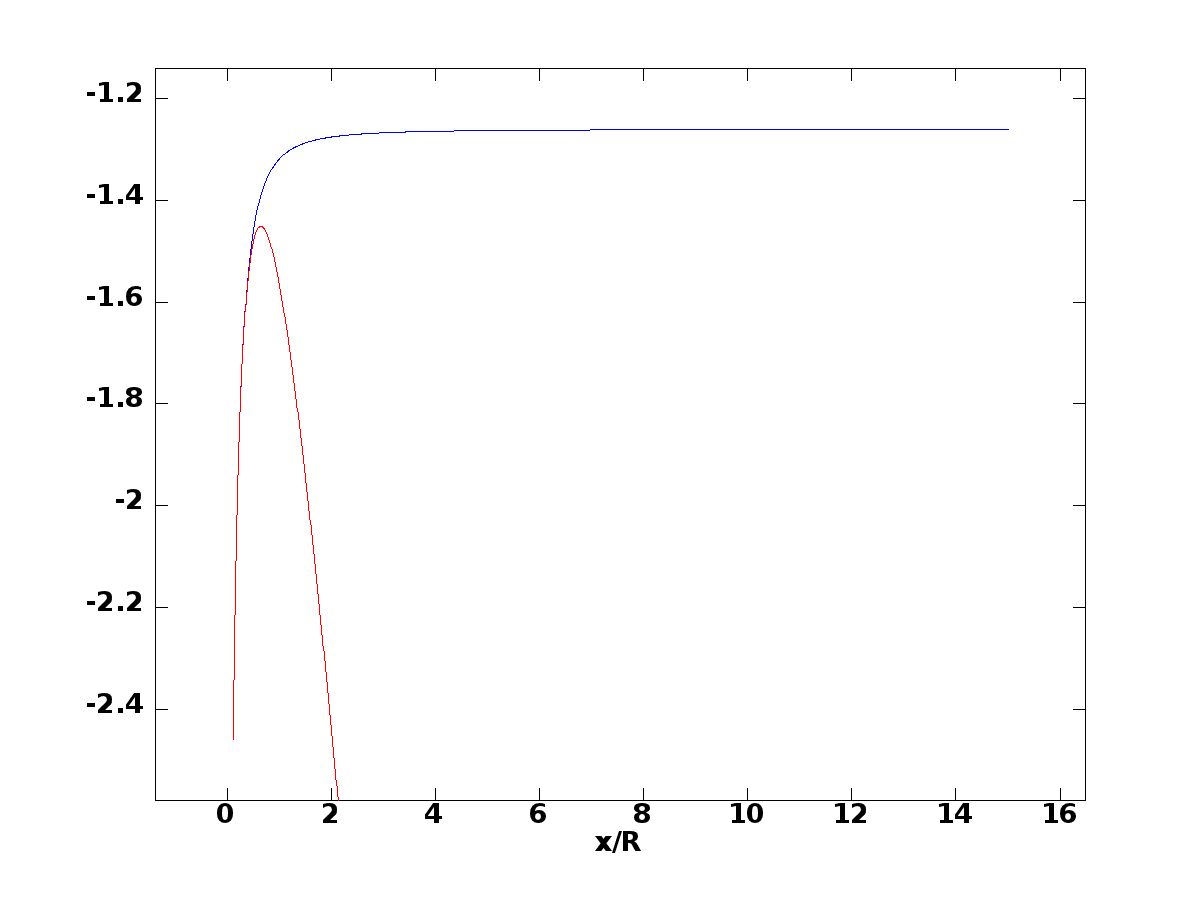}
\caption{the blue curve is the right side of the eq. (\ref{eq:GAP}), the red curve is with Q(x/R) omitted.}
\end{figure}

So let's consider the other case: $m=0, \sigma \neq 0$. Then (\ref{eq:begin}) reads
\begin{equation}
\frac{-g^2}{\pi R} \sum_{n=1} \int_{0}^{+\infty} \cfrac{dk}{k^2+(\cfrac{\pi n}{R})^2} -\frac{g^2 |\sigma|^2}{N} + 1 = 0
\end{equation}
Again using (\ref{eq:sum}), we obtain
\begin{equation}
\frac{g^2 |\sigma|^2}{N}=1-\frac{g^2}{2 \pi R} \int_0^{+\infty} dk \left( \frac{2R}{k} \left( \frac{1}{2} + 
\frac{1}{\exp(2Rk)-1} \right)  - \frac{1}{k^2}  \right) 
\end{equation}
Note that the integral is not divergent in infrared, as one might expect recalling the Mermin-Wagner-Coleman theorem. 
Indeed,
there is no spontaneous symmetry breaking at all: boundary conditions break $SU(N)$ to $SU(N-1)$ from the very beginning
and $SU(N-1)$ remains unbroken in all phases.
Due to Dirichlet boundary conditions we have natural IR cutoff $~ \pi/R$(see eq. (\ref{eq:ir_summ})). 
Using (\ref{eq:asympt})(if $m=0$ then $x=\infty$ ) we can write explicitly:
\begin{equation}
\frac{g^2 |\sigma|^2}{N}=1-\frac{g^2}{2 \pi} \left(\log(\La_{uv} R)+\gamma-\log(2 \pi) \right)
\end{equation}
or
\begin{equation}
\frac{|\sigma|^2}{N}=-\log(\La R)+\log(2 \pi)-\gamma
\end{equation}


\esec

\sec{Vacuum energy}
Above we have found the following effective action:
\begin{equation}
S_{eff}=iN Tr \log \l -\pr^2-\cfrac{\la g^2}{N} \r + \int{d^2 x \la}
\end{equation}
From now on, we will work in Euclidian space, so:
\begin{equation}
S_{eff,Eucl}=N Tr \log \l -\pr^2+\cfrac{\la g^2}{N} \r - \int{d^2 x \la}
\end{equation}
However, (\ref{eq:BOC}) breaks translation invariance and so $\langle0|T_{\mu \nu}|0 \rangle \neq \epsilon \eta_{\mu \nu}$, 
and
to calculate vacuum energy we will just calculate effective action. Using Pauli-Villars regularization(\cite{NOV}): 
\begin{equation}
\label{eq:reg}
S_{eff,Eucl}^{reg}=N \sum_{i=0}^2 c_i Tr \log \l -\pr^2+m^2+m_i^2 \r - \int{d^2x \la}
\end{equation}
\begin{equation}
m_0=0 \qs c_0=1 \qs c_1=\cfrac{m_2^2}{m_1^2-m_2^2} \qs c_2=\cfrac{-m_1^2}{m_1^2-m_2^2}
\end{equation}
At the end we should take limits $m_1 \rightarrow +\infty, m_2 \rightarrow +\infty$.

Regularized action should be stationary for $\la$ found above, so
\begin{equation}
\int{d^2x \cfrac{1}{g^2} }=\sum_{i=0}^2 c_i Tr \cfrac{1}{-\pr^2+m_i^2+m^2}
\end{equation}
Similar traces appeared above( eq. (\ref{eq:begin})) and they contained a nasty integral such as (\ref{eq:Q}). From now on,
we will consider the case $R \rightarrow +\infty$ in which calculation simplifies significantly. In this case 
(\ref{eq:approx}) is correct and nasty integral is of no interest due to $\exp(-2\sqrt{m_i^2+m^2}R)$ factor. 
After these remarks 
trivial calculation yields:
\begin{eqnarray}
\cfrac{1}{g^2}=\cfrac{1}{2 \pi R} \bigl( \cfrac{R}{2} \log(\cfrac{m^2+m_2^2}{m^2}) + \cfrac{R m_2^2}{2(m_1^2-m_2^2)} \
\log(\cfrac{m^2+m_2^2}{m^2+m_1^2}) - \cfrac{\pi}{2m} \nonumber \\
- \cfrac{\pi m_2^2}{2(m_1^2-m_2^2)} \cfrac{1}{\sqrt{m^2+m_1^2}} \
+ \cfrac{\pi m_1^2}{2(m_1^2-m_2^2)} \cfrac{1}{\sqrt{m^2+m_2^2}} \bigr)
\end{eqnarray}


Setting $m_1^2=xM^2,\; m_2^2=M^2$ and taking
\begin{equation}
\label{eq:limit}
x \rightarrow 1,\; M \rightarrow +\infty
\end{equation}
we obtain:
\begin{equation}
\label{eq:CC}
\cfrac{1}{g^2}=\cfrac{1}{2 \pi R} \l -\frac{R}{2} - \cfrac{\pi}{2m} \r =- \frac{1}{4 \pi} - \frac{1}{4mR}
\end{equation}

Regularized action (\ref{eq:reg}) contains $Tr \log(-\pr^2+m^2)$. It is well known that this is the Casimir 
energy for a massive complex  scalar field(\cite{CAS}). In 1+1:
\begin{equation}
\label{eq:Casimir}
E=-\frac{m}{2} -  \frac{R m^2}{\pi} \sum_{n=1}^{+\infty} \frac{K_1(2Rmn)}{Rmn}  
\end{equation}
where $K_1$ is modified Bessel function. 

The first term corresponds to the energy of boundary excitations. Usually it is omitted and the second term 
is called "the Casimir energy", but in our case $m$ depends on $R$, so 
the first term is important. If $mR>>1$, then the sum has the asymptotic behavior $\exp(-2mR)$ and so is negligible.

Expressions (\ref{eq:reg}), (\ref{eq:CC}) are free of divergences. $Trlog$ in (\ref{eq:reg}) could be calculated exactly via
Schwinger proper-time representation, but the expression is rather long and we will not give it here. After taking 
(\ref{eq:limit}) , we obtain $-\frac{Nm}{2}$ ($\exp(-2mR)$ term is dropped).
Therefore,
\begin{eqnarray}
E_{vac}=-\frac{Nm}{2}+\frac{NRm^2}{4 \pi} + \frac{Nm}{4} = \frac{NRm^2}{4 \pi}-\frac{Nm}{4}
\end{eqnarray}
where (\ref{eq:CC}) was used. There is no "interference" between two terms in (\ref{eq:reg}) and the limit (\ref{eq:limit})
can be taken separately.

Note that there is no mass parameter in the original Lagrangian. The mass is dynamically generated.
Therefore, to study R dependence in full we should take into account that $m$ depends on R. We 
will return to this fact in the next section.
Substituting (\ref{eq:m2}),(\ref{eq:m}), we arrive at
\begin{equation}
E_{vac}=\frac{N m_0^2 R}{4 \pi} - \frac{m_0 N}{2} + \frac{N \pi}{8 R}+O(1/R^2),\; R \rightarrow +\infty
\end{equation}

\esec
\sec{Discussion}
In \cite{SHIF} Shifman and Yung argued that for the $\mathbb{C}P(N-1)$ sigma model the L\"uscher coefficient follows
rich pattern of behavior, equals to $\cfrac{\pi N}{12}$ when $R<<\La^{-1}$ because $n^i$ could be considered massless,   
and approaches value of $0$ because $n^i$ are massive when $R>>\La^{-1}$. Indeed, we have seen that there is phase 
transition when $R \sim 1/\La$($R_{crit}=\exp(\log(2 \pi)-\gamma)/\La$ to be precise) and below this value $n^i$ are
massless. But above $1/\La$ we explicitly see Casimir-type behavior despite the existence of the mass gap.

However, in this situation the mass depends on $R$ and the L\"uscher term comes not from modified Bessel function in
(\ref{eq:Casimir})(as in the massless case) but from the first term which is
often of no physical meaning  but not in this case.
The considerations above led us to $-\cfrac{\pi N}{8}$ when $R>>\La^{-1}$. Note that the sign is opposite to one in usual 
Casimir energy expression \cite{CAS}.

In recent works \cite{INST_QCD}, \cite{CAS_QCD} Thomas and Zhitnitsky studied deformed QCD \cite{DEF_QCD} on 
$S^1 \times S^3$. By means of the monopole gas and the Sine-Gordon representations they argued that despite the existence
of the mass gap the vacuum energy obeys Casimir-type behavior $\sim 1/\mathbb{L}$($\mathbb{L}$ is the radius of 3-sphere) 
also with opposite sign. They relate it with the fact that the mass is not present in the theory from the very beginning, 
but emerges as a result of some dynamics. Obviously, it is the case of the $\mathbb{C}P(N-1)$ model.
\esec

\sec*{Acknowledgements}
Author is indebted to A. S. Gorsky for suggesting
this problem and numerous fruitful discussions.
\esec

\end{document}